\documentclass[twocolumn,aps]{revtex4-2}
\usepackage{amsmath}
\usepackage{amssymb}
\usepackage{graphicx}
\usepackage{dcolumn}
\usepackage{bm}
\usepackage{xcolor}
\usepackage{epstopdf}
\usepackage{longtable}
\usepackage{hyperref}

\begin{document}

\preprint{Phys.Rev.B }

\title{Charge transport in two-dimensional  conductors with hybrid three-component plasma}
\author{A. D. Levin,$^1$ V. A. Chitta,$^1$   Z. D. Kvon,$^{2,3}$ N. N. Mikhailov$^{2,3}$ and G. M. Gusev$^1$}

\affiliation{$^1$Instituto de F\'{\i}sica da Universidade de S\~ao
Paulo, 135960-170, S\~ao Paulo, SP, Brazil}
\affiliation{$^2$Institute of Semiconductor Physics, Novosibirsk
630090, Russia}
\affiliation{$^3$Novosibirsk State University, Novosibirsk 630090,
Russia}

\begin{abstract}
Electron--hole plasma in two-dimensional solid-state systems has previously been
studied mainly in two limiting cases: the degenerate regime in semimetals with
parabolic bands, and the nondegenerate symmetric Dirac regime in graphene-like
materials. Here we investigate a third, qualitatively different case realized in a
gapless HgTe quantum well: a multicomponent plasma with a hybrid
spectrum, in which massless Dirac carriers coexist with thermally activated heavy
holes from lateral valence-band valleys. Near the charge-neutrality point we observe
a pronounced increase of the resistance with temperature, following an approximately
$T^2$ law, in sharp contrast to the nearly temperature-independent resistivity
expected for a symmetric two-component Dirac plasma. We show that
this behavior originates in the charge-neutrality condition, which, once the heavy
holes are populated, pins the chemical potential above the Dirac point: the Dirac
electrons remain moderately degenerate while the heavy holes stay non-degenerate and
obey Boltzmann statistics. Because the light degenerate electrons scatter almost
elastically off the heavy Boltzmann holes, the interparticle transport admits a
relaxation-time description and can be treated non-perturbatively. Using carrier
densities from self-consistent band-structure calculations, we find that the data are
well reproduced by a short-range  interaction between Dirac
electrons and heavy holes in the weakly disordered limit, for which the electronic
conductivity scales as $T^{-2}$ and the excess resistivity is proportional to the
heavy-hole density, whereas an unscreened Coulomb interaction would leave the
conductivity temperature independent. The extracted interaction amplitude is of the
Coulomb scale expected at the low charge-neutrality density. Our results establish
near-critical HgTe quantum wells as a versatile platform for studying interaction-driven
transport in multicomponent systems combining degenerate massless and non-degenerate massive carriers, in which the linear Dirac spectrum decouples the current from the total momentum.
\end{abstract}
\maketitle

\section*{Introduction}

Charge transport in which the electric current is relaxed directly
by carrier--carrier collisions, rather than by disorder or phonons, is an unusual
regime in solid-state systems. This changes when the electric current is no longer
locked to a conserved quantity --- that is, when Galilean invariance is lifted so
that the current is not protected by total-momentum conservation. Interparticle
collisions may then relax the current directly, and their role becomes especially
pronounced at elevated temperatures. This non-invariance can arise in several distinct
ways: Fermi-surface geometries that allow Umklapp processes~\cite{pal};
electron--hole compensation in graphene and two-dimensional
semimetals~\cite{nam,tan,olshanetsky,entin}; multiband systems in which different
carrier groups have unequal effective masses and relaxation
times~\cite{hwang,nagaev1,nagaev2,gusev,levin}; and two-dimensional conductors with a
linear spectrum, where the relation between current and momentum differs from that in
parabolic bands~\cite{kovalev}. In the present system the Fermi surface near the
charge-neutrality point is small, so Umklapp processes are kinematically forbidden and
a parabolic spectrum would remain effectively Galilean-invariant; the non-invariance
relevant here is spectral in origin, arising from the linear Dirac dispersion, for
which the carrier velocity is not proportional to its momentum, so that the total
current is not proportional to the total momentum~\cite{kovalev}.

Two-dimensional semimetals, where electrons and holes coexist near charge neutrality,
constitute a particularly well-studied class of such systems. In the degenerate
regime, the phase space for interparticle scattering produces a characteristic
temperature dependence of the resistivity; a $T^2$ and $T$ resistivity driven by
electron--hole collisions has been observed in wide HgTe quantum
wells~\cite{entin,snegirev}. Dirac materials offer a qualitatively different
realization of the electron--hole plasma. Because of the linear spectrum and the
vanishing density of states at the charge-neutrality point, both the carrier density
and the effective inertial mass of thermally excited carriers are strongly temperature
dependent. The most prominent solid-state example is graphene, where high-quality
samples and elevated temperatures are required to overcome disorder-induced
electron--hole puddles~\cite{sarma,xin}. In this regime, graphene displays hallmarks
of strong electron--hole interactions: hydrodynamic flow, violation of the
Wiedemann--Franz law~\cite{crossno2016}, giant magnetoresistance, and Coulomb
drag~\cite{nam,tan,xin,ponomarenko}. For a symmetric two-component Dirac plasma,
however, the thermally excited carrier density ($\propto T^2$) and the effective Dirac
mass ($\propto T$) compensate each other when scattering follows the Planckian form
$1/\tau_{eh}\sim k_BT/h$~\cite{phillips}, yielding an approximately
temperature-independent resistivity near the charge-neutrality point~\cite{xin} --- a
useful benchmark for identifying deviations from the symmetric two-component picture.

Gapless HgTe quantum wells provide a unique opportunity to explore such deviations.
Near the critical well thickness, HgTe hosts massless Dirac fermions with a linear
spectrum similar to graphene, but the valence band additionally contains lateral
heavy-hole valleys below the Dirac point. These massive states carry
a large density of states and become thermally populated at elevated temperatures,
realizing a multicomponent plasma that is absent in graphene. 

Crucially, the large
heavy-hole density of states has a qualitative consequence at charge neutrality: once
the heavy holes are populated, the neutrality condition forces the chemical potential
above the Dirac point, so that the Dirac electrons become moderately degenerate while
the heavy holes remain non-degenerate and Boltzmann-distributed. The system is thus a
hybrid of degenerate massless and non-degenerate massive carriers --- distinct both
from the degenerate parabolic semimetals and from the symmetric non-degenerate Dirac
plasma of graphene.

Here we study charge transport near the charge-neutrality point in
this system. Experimentally, the resistance increases with temperature and follows an
approximately $T^2$ law above $\approx10$~K, in contrast to the nearly
temperature-independent resistivity of a symmetric Dirac plasma. We show that this
behavior reflects the temperature evolution of the carrier composition, which drives a
crossover from a low-temperature symmetric two-component Dirac plasma to a
high-temperature hybrid plasma of degenerate Dirac electrons and Boltzmann heavy holes;
the intermediate three-component window, in which Dirac electrons, Dirac holes, and
heavy holes contribute simultaneously, is narrow for the broadening realized in our
samples and is not resolved separately. 

Because the light degenerate electrons scatter
almost elastically off the heavy Boltzmann holes, the interparticle transport admits a
momentum-relaxation-time description and can be treated non-perturbatively, following
the kinetic theory of Ref.~\cite{kovalev2}. In this framework an effective short-range electron--heavy-hole interaction in the weakly disordered limit yields an
electronic conductivity $\sigma_e\propto T^{-2}$, and hence the observed quadratic
resistivity, whereas an unscreened Coulomb interaction would leave the conductivity
temperature independent.
We further find that the disorder inherent to these near-critical
wells plays a constructive role. The broadening of the heavy-hole band edge by
well-width fluctuations --- the same disorder that produces the residual percolation
conductivity near charge neutrality (Sec.~III) --- populates the heavy holes already
at low temperatures, bringing the degenerate hybrid regime, which for an ideal sharp
gap would appear only above $\approx50$~K, down into the experimentally accessible
window. Using carrier densities from self-consistent band-structure calculations, we
show that the excess resistivity is proportional to the heavy-hole density and that the
extracted short-range amplitude is of the Coulomb scale expected at the low
charge-neutrality density.

\section {Energy spectrum of the Dirac fermions in HgTe quantum wells}
\begin{figure*}[ht]
\includegraphics[width=18cm]{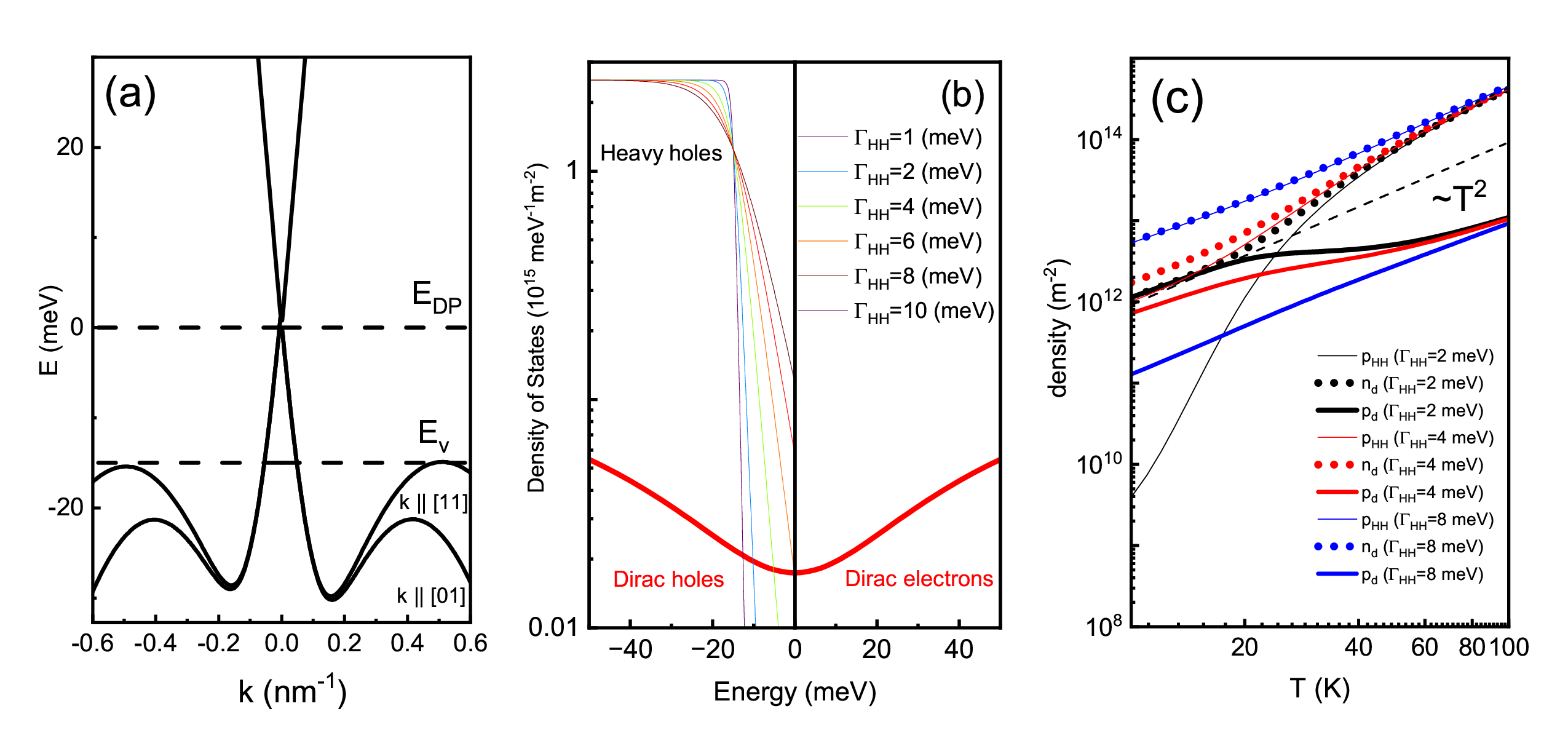}
\caption{\label{fig1}(Color online)
a) Representation of the energy spectrum of a 6.3-nm mercury telluride quantum well. (b) Density of states of the Dirac carriers  and heavy holes
for different values of the heavy-hole broadening parameter $\Gamma_{HH}$.
(c) The carrier density for different branches as a function of temperature
for different heavy-hole broadening parameter
$\Gamma_{HH}$. $\mu$ is fixed self-consistently by charge neutrality. Dashes indicate a $T^2$ dependence for comparison.}
\end{figure*}
\begin{figure}
\includegraphics[width=9cm]{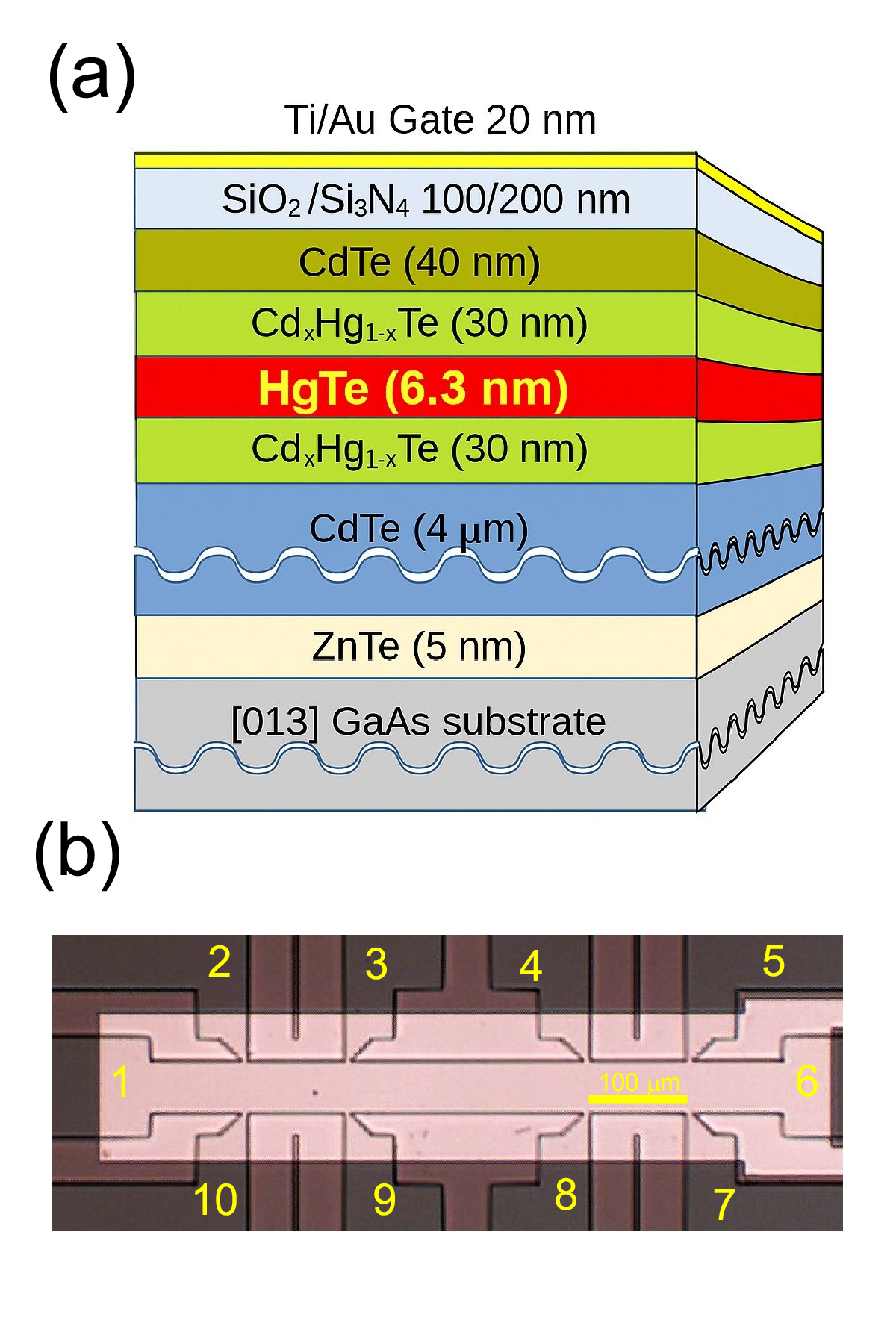}
\caption{\label{fig2}(a) Schematic illustration of the gated HgTe quantum-well
transistor structure. (b) Representative top view of the Hall-bar device.
The numbered regions denote ohmic contacts; the central channel, voltage
probes, and gated area are indicated. Both samples share the same nominal
geometry. The scale bar is \(100~\mu\mathrm{m}\).}
\label{sample}
\end{figure}
 
\begin{figure*}[ht]
\includegraphics[width=18cm]{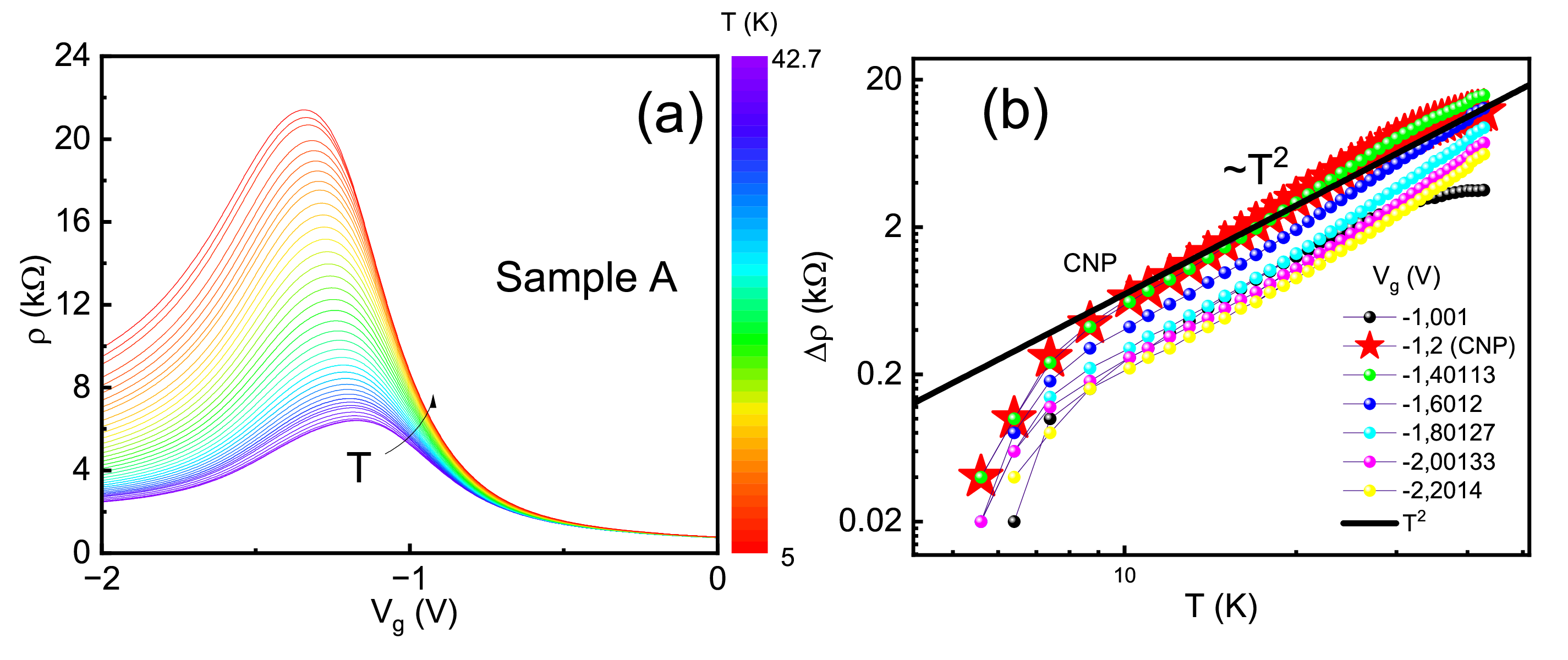}
\caption{\label{fig3}(Color online)
(a) Resistivity as a function of the gate voltage at different temperatures for HgTe gapless quantum well ( sample A). (b) Excess resistivity $\rho (T)-\rho_0$ as a function of the temperature for different gate voltages, where $\rho_0$ is the extrapolated zero-temperature background (Sec. III). The star symbols mark the trace at the fixed low-temperature
charge-neutrality reference voltage $V_0$.  Solid line shows $T^2$ dependence for comparison.}
\end{figure*}

\begin{figure*}[ht]
\includegraphics[width=18cm]{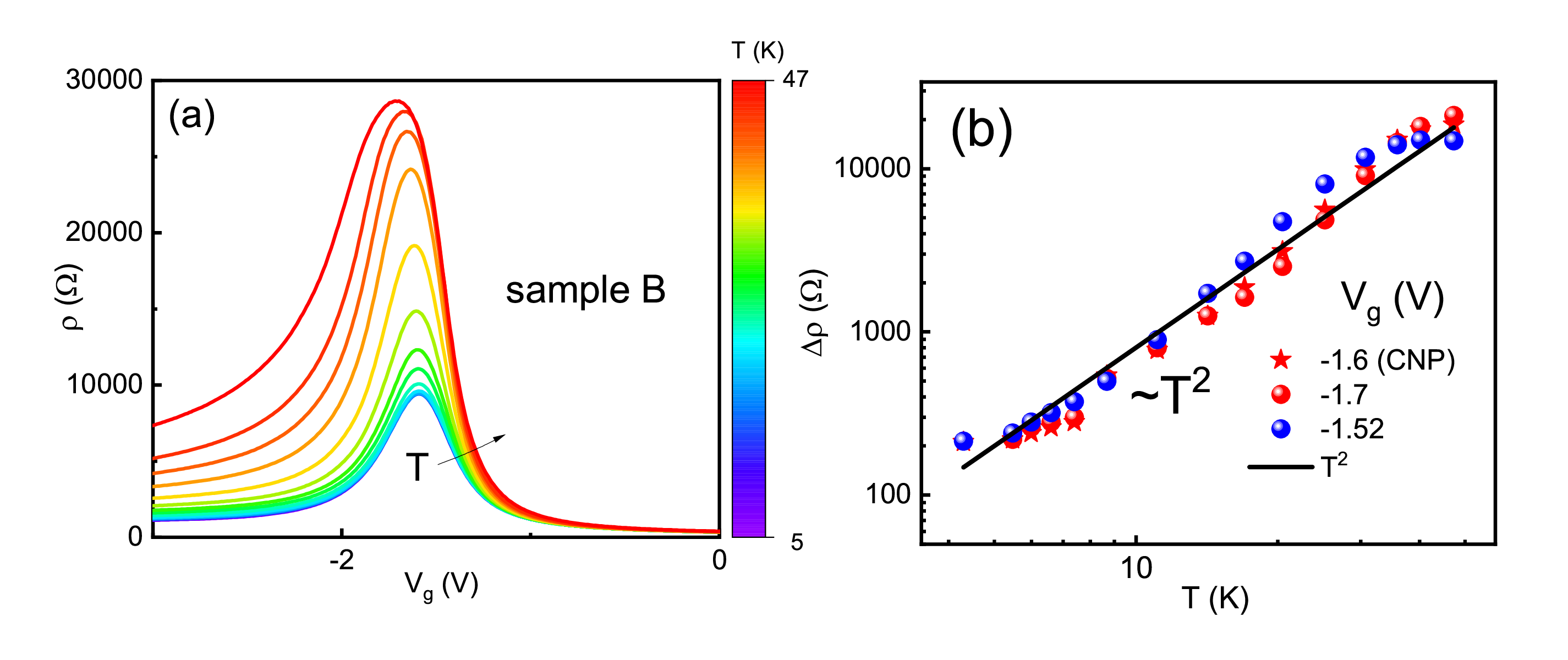}
\caption{\label{fig4}(Color online)
(a) Resistivity as a function of the gate voltage at different temperatures for HgTe gapless quantum well (sample B). (b) Excess resistivity $\rho (T)-\rho_0$ as a function of the temperature for different gate voltages, where $\rho_0$ is the extrapolated zero-temperature background (Sec. III). The star symbols mark the trace at the fixed low-temperature
charge-neutrality reference voltage $V_0$. Solid line shows $T^2$ dependence for comparison.}
\end{figure*}
To investigate the transport properties of charge carriers in a gapless HgTe quantum well, we begin by presenting the energy spectrum across a broad interval for both the conduction and valence bands, as shown in Fig.~\ref{fig1}(a). This system hosts Dirac fermions with a linear dispersion relation for electrons
and holes, given by $\varepsilon_e=\pm \hbar v|k|$, where $v=7\times10^{7}$~cm/s
$=c/430$. The band-structure parameters used here --- the Dirac velocity
$v_F$, the heavy-hole offset $\Delta=|E_v-E_{DP}|$, and the heavy-hole effective
mass $m_{HH}$ --- are taken from eight-band and six-band $k\cdot p$ calculations for
HgTe quantum wells of this type~\cite{raichev2012, kristopenko2016} and
from experiment. Figure~1(a) shows the corresponding spectrum: the linear Dirac
cone of the near-critical gapless well and the lateral heavy-hole maxima at
$E_v=-\Delta$, with the slope and the positions of the maxima fixed by these
parameters. All energies are measured from the Dirac point,
$E_{DP}=0$. A lateral maximum is observed in the valence band below the Dirac
point, at $E_v=-\Delta$, where $\Delta=|E_v-E_{DP}|\approx 15$~meV is a fixed
band-structure offset; the momentum position and energy of this lateral maximum
are properties of the band structure and are not set by the chemical potential
$\mu$. A degenerate heavy-hole Fermi contour, with radius fixed by
$E_{HH}(q)=\mu$, would exist only when the chemical potential lies below the
heavy-hole maximum, $\mu<E_v$ (i.e.\ deeper into the valence band). Near the
charge-neutrality point, however, $\mu\approx 0$ lies above $E_v=-\Delta$, so the
heavy holes are thermally activated rather than characterized by a zero-temperature
Fermi momentum. The heavy-hole density is therefore obtained from the broadened
density of states and the Fermi distribution (Sec.~I), rather than from a
zero-temperature momentum expression.

The degeneracy factors used in the density calculation are defined here.
For the single Dirac cone we take $g_d=2$, corresponding to spin degeneracy. For the heavy-hole side valleys we take $g_{HH}=g_s g_v=4$, with spin degeneracy $g_s=2$ and valley degeneracy $g_v=2$. The heavy-hole effective mass is \(m_{HH}=0.15m_0\).
These factors are included directly in the densities of states used below.
Consequently, the carrier densities \(n_d\), \(p_d\), and \(p_{HH}\) denote
total sheet densities, not per-flavor densities. This band structure has been
extensively studied in prior works~\cite{buttner, kozlov2013, gusev2, kristopenko2019},
and our calculations follow the model detailed in Ref.~\cite{gusev2}.
Figure~1(b) displays  the density of states (DOS) for Dirac carriers and heavy holes for varying heavy-hole broadening parameters  $\Gamma_{HH}$. The DOS is shown
on a logarithmic scale, so that the small Dirac DOS and the much larger heavy-hole
DOS are both visible without rescaling. Due to their large effective mass and
valley degeneracy ($g_v=2$), the DOS of heavy holes (HH) significantly exceeds that
of Dirac holes (DH).
In practical samples, unavoidable fluctuations in the quantum well (QW) width
during growth introduce in-plane variations in the bandgap~\cite{gusev2022}. These
variations affect the positions of the DP and $E_v$, leading to a broadening of the
DOS, as illustrated in Fig.~1(b). Based on the topological network model from
Ref.~\cite{gusev2022}, we estimate this broadening to be in the range of
1--4~meV. We note, however, that the value entering the
high-temperature transport fit (Sec.~V), $\Gamma_{HH}\approx 8$~meV, is larger than
this estimate and is constrained by the transport data rather than determined
independently; this is discussed further in Sec.~V.
The heavy-hole DOS is obtained by convolving the sharp parabolic-band DOS
with a Gaussian distribution of local heavy-hole band-edge energies of width
\begin{widetext}
$\Gamma_{HH}$:
\[
D_{HH}^{\mathrm{conv}}(E)=
\int_{-\infty}^{\infty} dE_v'\,
\frac{1}{\sqrt{2\pi}\Gamma_{HH}}
\exp\!\left[-\frac{(E_v'-E_v)^2}{2\Gamma_{HH}^2}\right]
\frac{g_{HH}m_{HH}}{2\pi\hbar^2}
\Theta(E_v'-E),
\]
\end{widetext}
where \(E_v=-\Delta\). This convolution produces a finite tail that can extend
formally above the Dirac point. Since the heavy-hole states belong to the
valence band, this positive-energy tail should not be interpreted literally.
A fully microscopic treatment would require a self-consistent calculation of
the broadened band edge and local electrostatic potential. In the present
phenomenological model we therefore impose a cutoff in the electron region,
\[
D_{HH}(E)=D_{HH}^{\mathrm{conv}}(E)\Theta(-E),
\]
i.e. \(D_{HH}(E>0)=0\). This cutoff is a simplifying approximation used to
avoid assigning heavy-hole spectral weight to the electron side of the Dirac
point.

The Dirac density of states vanishes linearly at the Dirac point
in the ideal case; in the presence of disorder it is rounded near $E=0$ by the same
spatial fluctuations of the quantum-well width that broaden the heavy-hole edge.

In the calculations below we use a temperature-independent gapless
spectrum as a minimal effective reference model, with the temperature
dependence entering through the carrier occupations. This assumption
should not be interpreted as an experimental demonstration that the
band structure remains strictly gapless at finite temperature. HgTe
quantum wells close to the critical thickness are particularly sensitive
to temperature: multiband calculations predict a substantial
temperature dependence of the critical well thickness and the possible
opening of a normal gap upon heating
\cite{yahniuk2019,kristopenko2019}, while temperature-induced changes
of the band structure have also been observed experimentally
\cite{Wiedmann2015,Marcinkiewicz2017,Kadykov2018}.

The absence of a simple activated temperature dependence does not, by
itself, exclude the presence of a small temperature-dependent gap.
In the present asymmetric multicomponent system, a finite gap would
modify not only the Dirac spectrum but also the self-consistent chemical
potential and the relative populations of Dirac electrons, Dirac holes,
and heavy holes. Its influence on the resistance therefore cannot be
identified with a simple activation law and would require a
self-consistent recalculation of both the carrier populations and the
transport coefficients.

As an order-of-magnitude estimate, an eight-band Kane calculation for a
nominal 6.3-nm [013] HgTe quantum well gives a direct $\Gamma$-point gap
of approximately $0.9$, $3.4$, and $6.1$~meV at 10, 20, and 30~K,
respectively. Because such small gaps near the critical thickness are
highly sensitive to the precise well width, strain, band offsets, and
temperature-dependent band parameters, we do not use these values as an
additional fitting constraint.

Accordingly, the present zero-field transport data do not provide a
unique quantitative determination of $E_g(T)$ and cannot exclude a small
temperature-induced gap. The gapless spectrum used below is therefore
treated as a minimal effective reference model for isolating the effect
of the evolving carrier composition, rather than as a band structure
demonstrated experimentally to remain strictly gapless at finite
temperature.

For this reference spectrum, the finite-temperature carrier populations
are calculated directly from the corresponding densities of states.
We denote by $n_d$ the total density of Dirac electrons, by $p_d$ the
total density of Dirac holes, and by $p_{HH}$ the total density of
heavy holes. The density of states $D_d(\varepsilon)$ includes the
degeneracy of the Dirac cone, while $D_{HH}(\varepsilon)$ includes the
spin and valley degeneracies of the heavy-hole valleys. No additional
degeneracy factors are therefore introduced in the carrier densities,
which are given by

\begin{equation}
n_d(T)=\int_0^\infty d\varepsilon\,
D_d(\varepsilon) f(\varepsilon-\mu),
\end{equation}

\begin{equation}
p_d(T)=\int_{-\infty}^0 d\varepsilon\,
D_d(\varepsilon)\left[1-f(\varepsilon-\mu)\right],
\end{equation}

\begin{equation}
p_{HH}(T)=\int_{-\infty}^0 d\varepsilon\,
D_{HH}(\varepsilon)\left[1-f(\varepsilon-\mu)\right],
\end{equation}

where

\begin{equation}
f(\varepsilon-\mu)=
\left[
1+\exp\left(\frac{\varepsilon-\mu}{k_BT}\right)
\right]^{-1}.
\end{equation}

Here $D_d(\varepsilon)$ is the total Dirac density of states,
$D_{HH}(\varepsilon)$ is the broadened heavy-hole density of states,
and $\mu$ is the chemical potential.

The gate voltage fixes the net sheet charge rather than the individual
carrier populations. At the charge-neutrality point, the chemical
potential is therefore determined self-consistently from

\begin{equation}
n_d-p_d-p_{HH}=0.
\label{eq:neutrality_g}
\end{equation}

Away from charge neutrality, the right-hand side is replaced by the
corresponding gate-induced density. Thus, as temperature changes,
$n_d$, $p_d$, $p_{HH}$, and $\mu$ evolve together while satisfying the
gate-controlled charge balance.

Figure~\ref{fig1}(c) illustrates this evolution for different values of
the heavy-hole broadening $\Gamma_{HH}$. At sufficiently low
temperature the plasma is dominated by thermally excited Dirac
electrons and Dirac holes, and the chemical potential remains close to
the Dirac point. As the temperature increases, heavy holes from the
lateral valence-band valleys become populated. Their large density of
states has an important consequence: charge neutrality shifts the
chemical potential above the Dirac point, increasing the Dirac-electron
population while suppressing the relative contribution of Dirac holes.

The temperature at which this crossover occurs is controlled strongly
by the heavy-hole broadening. Increasing $\Gamma_{HH}$ extends the
heavy-hole density of states toward the Dirac point and therefore moves
the onset of heavy-hole population to lower temperatures. For the
broadening relevant to the experimental samples, the heavy-hole
contribution is already appreciable around $T\sim10$~K, and the
Dirac-hole density rapidly becomes subdominant.

The system therefore passes through a narrow three-component crossover
regime in which Dirac electrons, Dirac holes, and heavy holes coexist.
At higher temperatures, however, charge compensation is dominated by
Dirac electrons and heavy holes, and the plasma can be treated to a
good approximation as an effective two-component hybrid system of light
Dirac electrons and massive holes. In this regime the calculated
Dirac-electron and heavy-hole densities exhibit an approximately
quadratic temperature dependence over the experimentally relevant
interval, as discussed quantitatively below.
\section{Methods}
We fabricated quantum wells using HgTe/Cd$_x$Hg$_{1-x}$Te heterostructures grown on substrates with a $[013]$ crystallographic orientation. The two quantum wells had nominal thicknesses of
$d=6.3$ and $6.4$~nm, respectively. During the molecular beam epitaxy (MBE) growth, the layer thickness was continuously monitored via in situ ellipsometry, achieving a precision of $\pm 0.3$~nm (Fig.~\ref{sample}a).

The fabricated devices consisted of multiterminal Hall bar structures. The  configuration comprised a uniform-width Hall bar segmented into three consecutive sections, each 50~$\mu$m wide and with lengths of 100~$\mu$m, 250~$\mu$m, and 100~$\mu$m, respectively (Fig.~\ref{sample}b). 

Electrical contacts were defined using standard lithographic techniques and formed by thermally bonding indium to the designated contact pads. Due to the low MBE growth temperature (approximately $180^{\circ}$C), the subsequent contact fabrication was also carried out at low temperatures. The indium diffused vertically into the contact regions, forming ohmic contacts to the quantum well. The resulting contact resistance varied between 10~k$\Omega$ and 50~k$\Omega$.

Throughout all AC measurements, the reactive component of the total impedance was maintained below 5\%, confirming the quality of the ohmic contacts. Furthermore, the current-voltage ($I$-$V$) characteristics exhibited linear behavior at low bias, indicating ohmic conduction. For Device A, the current $I$ was applied between contacts 1 and 6, and the voltage $V$ was measured across probes 3 and 4, resulting in a four-terminal resistance $R = R_{1,6}^{3,4} = \frac{V_{3,4}}{I_{1,6}}$. The measured four-terminal resistance \(R\) was converted to the two-dimensional sheet resistivity as
$\rho(T)=\frac{W}{L}R(T)$, where \(W\) is the Hall-bar width and \(L\) is the distance between the voltage probes.

A 200~nm-thick SiO$_2$ dielectric layer was deposited on  top of the sample, followed by a Ti/Au metallic gate. Based on the dielectric thickness and Hall measurements, the gate-induced carrier density modulation was estimated to be approximately $0.9 \times 10^{11}$~cm$^{-2}$/V, consistent with prior studies employing similar device architectures.

Their key parameters for two samples, labeled A and B, fabricated on different substrates are summarized in Table I, including the quantum well thickness ($d$), the gate voltage corresponding to the the low-temperature charge-neutrality reference voltage ($V_0$), the maximum resistivity at the Dirac point ($\rho_{\mathrm{max}}$), and the electron mobility ($\mu_e$), calculated using $\mu_e = 1/\rho eN_s$ with a nominal carrier density of $N_s = 1 \times 10^{11}$~cm$^{-2}$.

\begin{table}[ht]
\centering
\begin{tabular}{|l|c|c|c|c|}
\hline
Sample & $d$ (nm) & $V_{\mathrm{0}}$ (V) & $\rho_{\mathrm{max}}$ ($h/e^2$) & $\mu_e$ (cm$^2$/V$\cdot$s) \\
\hline
A & 6.3 & $-1.2$ & 0.25 & 85.000 \\
\hline
B & 6.4 & $-1.6$  & 0.36 & 104.800 \\
\hline
\end{tabular}
\caption{Representative parameters of the HgTe quantum-well samples
measured at $T=4.2$~K.}
\label{tab:sample_parameters}
\end{table}
\section{Experiment in reference device}
The resistance behavior in gapless semimetals, including graphene-like systems, typically exhibits a characteristic dependence on gate voltage: a pronounced resistivity peak aligned with the charge neutrality point. In Dirac semimetals, this point coincides with the Dirac point at zero energy. In our HgTe quantum well samples, we observed a similar gate-dependent behavior in both devices, as illustrated in Figs.~\ref{fig3} and~\ref{fig4}. 

However, notable deviations from monolayer graphene emerge near the Dirac point. Specifically, spatial fluctuations in the quantum well (QW) thickness around the average critical width $d = d_c$, which are inherently present due to limitations in the growth process, lead to local variations in the energy gap. These variations can induce random sign changes in the gap, resulting in transitions between topologically distinct phases—namely, the normal insulator and the two-dimensional topological insulator (2D TI) phases. Consequently, a complex network of zero-energy modes forms along the interfaces separating these regions.

In addition to gap fluctuations, disorder in the form of spatially varying electrostatic potential—originating from randomly distributed charged impurities—further modulates the local band structure. When the Fermi level is tuned near the Dirac point, the electrical conductivity of the system is dominated by percolation through these zero-energy channels.

Previous transport measurements in similar near-critical HgTe quantum wells
found conductivities in the range $(1.5-4)e^2/h$, consistent with the
topological-network picture of Ref.~\cite{gusev2022}. In this study, we focused on the temperature dependence of the resistivity near the charge neutrality point (CNP).  Figures~\ref{fig3} and~\ref{fig4} present the evolution of resistivity as a function
of temperature for both samples. A similar trend is observed in both cases: the
resistivity exhibits a pronounced increase with rising temperature in the vicinity of
the peak, while the position of the resistivity maximum shifts toward more negative
gate voltage as the temperature increases.

Two distinct quantities should be distinguished here. We denote by
$V_0$ the gate voltage corresponding to the charge-neutrality point
identified at low temperature. This voltage is reproducible upon repeated
warming and cooling cycles, with no detectable hysteresis. We therefore
use $V_0$ as a fixed experimental reference throughout the
temperature-dependent analysis.

At a fixed gate voltage, the net gate-induced carrier density is taken
to remain fixed, whereas its distribution among Dirac electrons,
Dirac holes, and heavy holes changes with temperature. Thus,
\[
n_d-p_d-p_{\rm HH}
=
\alpha_g(V_g-V_0),
\]
while $n_d$, $p_d$, $p_{\rm HH}$, and the chemical potential
$\mu$ evolve with temperature.

Additional support for using a fixed electrostatic reference comes from
magnetotransport measurements on near-critical HgTe quantum wells.
Although a Hall analysis directly at the CNP is complicated by the
simultaneous presence of several carrier components, the situation is
considerably simpler on the electron side, where transport is predominantly
single-carrier and the Hall coefficient and Shubnikov--de Haas oscillations
provide direct constraints on the electron density. In particular, well-developed quantum Hall features were observed
in a 6.6-nm near-critical HgTe quantum well at $T=80$~K, with
magnetotransport measurements performed over a broad temperature range
\cite{kozlov2014}. Together with previous observations that the Hall
response and Shubnikov--de Haas oscillations show no substantial changes
below approximately $70$~K, this strongly constrains any large
temperature-induced change of the net electrostatic carrier density in
the temperature range relevant here.

The position $V_{\rm peak}(T)$ of the resistance maximum is a different
quantity. In a multicomponent system with unequal carrier mobilities,
the minimum of
\[
\sigma(V_g,T)
=
e\left[
n_d\mu_e+p_d\mu_d+p_{\rm HH}\mu_{\rm HH}
\right]
\]
does not, in general, coincide exactly with charge neutrality.
Consequently, a temperature-induced shift of the resistance maximum
does not by itself imply a corresponding shift of the electrostatic
charge-neutrality voltage.

To estimate this effect quantitatively, we performed the simplified
calculation described in the Supplemental Material ~\cite{Supplemental}. The effective
mobilities are determined from the experimental 5-K gate-voltage
dependence and then kept fixed, while only the carrier populations are
recalculated at 42.7 K. Experimentally, the resistance maximum shifts by
approximately
$\Delta V_{\rm peak}^{\rm exp}\simeq -171~{\rm mV}$,
whereas the calculated redistribution of the three carrier populations
at fixed $V_0$ produces
$\Delta V_{\rm peak}^{\rm calc}\simeq -54~{\rm mV}$.
Thus, the change in carrier composition alone produces a shift in the
observed direction and accounts for approximately 32\% of the measured
displacement.

The remaining quantitative difference should not be interpreted as
evidence for a temperature-dependent $V_0$. Within the present
zero-field measurements, it cannot be separated from the temperature
and gate-voltage dependence of the individual mobilities and
intercomponent scattering rates. A further separation of these effects
would require Hall and magnetoresistance measurements.

For the quantitative temperature analysis below, we therefore follow
the fixed low-temperature reference voltage $V_0$, rather than the
moving resistance maximum, and define
\[
\Delta\rho(T)=\rho(V_0,T)-\rho_0,
\]
where $\rho_0$ is the extrapolated zero-temperature background obtained
from the low-temperature fit
\[
\rho(T)=\rho_0+CT^2.
\]
We use this extrapolated background rather than the measured value
$\rho(5\,\mathrm{K})$, because a substantial quadratic contribution has
already developed by $5\,\mathrm{K}$ ($\approx220\,\Omega$ for sample A).
Subtracting $\rho(5\,\mathrm{K})$ would therefore remove part of the
$T^2$ contribution under study. The same $\rho_0$ enters as the
disorder-limited residual background in the two-parameter fit of
Sec.~IV [Eq.~(17)].

The two devices were not measured with equal accuracy, which is directly
visible in the data.

Sample~A was measured on a dense, uniform temperature
grid with little scatter and carries the quantitative analysis. A free-exponent fit
$\Delta\rho\propto T^{p}$ over $10$--$30$~K gives $p=2.12\pm0.02$ ($R^2=0.998$) for
sample~A; the local logarithmic slope, evaluated over the same interval, remains in
the range $2.1$--$2.3$ and decreases above $T\simeq30$~K, confirming that the
dependence is only approximately quadratic and contains a crossover at higher
temperature. The fit residuals for sample~A are small (rms $\sim6\%$, within
$\sim10\%$ across the interval). 

Sample~B was measured on a sparser, uneven grid and
shows substantially larger scatter; being the more resistive device
($\rho_{\max}=0.36$ vs $0.25\,h/e^2$) it is also more affected by self-heating, and
its low-temperature background is less well constrained. The same fit to sample~B
gives a nominally larger exponent ($p\approx2.6$) with an early high-temperature
saturation and residuals up to $\sim20\%$; in view of this larger, partly systematic
uncertainty we do not regard it as a reliable determination of the intrinsic
exponent, and sample~B serves as a qualitative confirmation that the resistance rises
strongly with temperature through heavy-hole population.

\section{Theory and discussion}

The analytical treatment of this section is formulated on the ideal
charge-neutrality trajectory,
\[
n_d-p_d-p_{\rm HH}=0.
\]
For comparison with experiment, we use the fixed low-temperature
reference voltage $V_0$ introduced in Sec.~III as an experimental proxy
for this trajectory. The quantitative comparison in Fig.~5 is therefore
performed at $V_0$, while the full gate-voltage sweeps and the off-reference
temperature traces shown in Figs.~3 and 4 are presented as additional
experimental data and are not described by the analytical formulas below.

According to the calculated carrier densities [Fig.~\ref{fig1}(c)],
the composition of the plasma near the charge-neutrality point (CNP) evolves
continuously with temperature, driving a crossover between two well-defined regimes.
At low temperatures, \(T\lesssim10\) K, the heavy-hole contribution is negligible and
the system is an approximately symmetric two-component Dirac plasma of Dirac electrons
and Dirac holes, with the chemical potential at the Dirac point ($\mu\simeq0$). As the
temperature increases, the heavy holes -- which carry a large density of states -- are
thermally populated and progressively replace the Dirac holes, so that at
\(T\gtrsim10\) K the system becomes a hybrid plasma dominated by Dirac electrons and
heavy holes. The intermediate window in which all three species (Dirac electrons,
Dirac holes, and heavy holes) contribute simultaneously is narrow for the broadening
realized in our samples and is not resolved as a separate regime; its extent is set by
the heavy-hole broadening \(\Gamma_{HH}\), stronger broadening shifting the onset of
the hybrid regime to lower temperatures.

A central feature of the hybrid regime, dictated by charge neutrality,
distinguishes it from both the symmetric Dirac plasma and the degenerate parabolic
semimetals. Because the heavy-hole density of states greatly exceeds that of the Dirac
carriers, maintaining neutrality once the heavy holes are populated forces the
chemical potential to rise above the Dirac point. As a result, the Dirac electrons
become \emph{moderately degenerate} (\(\mu/k_BT\simeq1.3\)--\(2.2\) over the measured
range), while the heavy holes remain \emph{deeply non-degenerate} and follow Boltzmann
statistics. 

We treat the two limiting regimes in turn: the low-temperature symmetric
Dirac plasma (Sec.~IV\,A) and the high-temperature degenerate
Dirac-electron--Boltzmann-hole plasma (Secs.~IV\,B and IV\,C), on which the
quantitative analysis of the experiment is based. 
The crossover can be characterized quantitatively by the fractional
heavy-hole contribution to charge neutrality. Since the degeneracy factors
are already included in the densities of states, the carrier densities
\(n_d\), \(p_d\), and \(p_{HH}\) denote total sheet densities. We therefore
define
\[
\eta_{HH}(T)=
\frac{p_{HH}(T)}{n_d(T)} .
\]
Using the charge-neutrality condition (Eq.~(5)), one has
\[
\eta_{HH}(T)+\eta_d(T)=1,
\qquad
\eta_d(T)=\frac{p_d(T)}{n_d(T)}.
\]

The same compensation can also be characterized by the ratio of
heavy-hole to Dirac-hole densities,

\[
R_{HH/d}(T)=\frac{p_{HH}(T)}{p_d(T)}
=
\frac{\eta_{HH}(T)}{1-\eta_{HH}(T)}.
\]
Thus the low-temperature symmetric Dirac-plasma regime corresponds to
\(\eta_{HH}\ll 1\), or \(R_{HH/d}\ll 1\), where the Dirac electrons are
compensated mainly by Dirac holes. The crossover occurs when
\[
p_{HH}\sim p_d
\quad
(R_{HH/d}\sim 1,\ \eta_{HH}\sim 1/2),
\]
and the high-temperature hybrid regime is reached when
\[
\eta_{HH}\to 1
\quad
(R_{HH/d}\gg 1),
\]
so that Dirac holes are subdominant and charge neutrality is maintained
mainly by Dirac electrons and thermally activated heavy holes. In practice
we use \(R_{HH/d}>1\) as the onset of heavy-hole domination and
\(\eta_{HH}\gtrsim 0.8\) as the regime where the two-component
Dirac-electron--heavy-hole approximation is quantitatively meaningful.

\subsection{Low temperature: symmetric two-component Dirac plasma}

Before considering the high-temperature hybrid regime, we first discuss
the low-temperature limiting case of the full three-component system.
When the heavy-hole population is negligible,
$p_{\rm HH}\ll p_d$, the charge-neutrality condition reduces to
$n_d\simeq p_d$, and the system approaches a symmetric two-component
plasma of Dirac electrons and Dirac holes. This subsection is introduced
as an auxiliary reference limit, providing a benchmark for the subsequent
crossover to the Dirac-electron--heavy-hole regime considered in
Secs.~IV.B and IV.C.

In the presence of an electric field
\(\mathbf{E}=E\hat{\mathbf{x}}\), the coupled equations for the drift
velocities \(u_e\) and \(u_h\) are
\begin{equation}
\left\{
\begin{aligned}
-\frac{e}{m^*(T)}E
&=
\frac{1}{2\tau_{eh}(T)}(u_e-u_h)+\frac{u_e}{\tau},
\\[0.4em]
\frac{e}{m^*(T)}E
&=
-\frac{1}{2\tau_{eh}(T)}(u_e-u_h)+\frac{u_h}{\tau},
\end{aligned}
\right.
\label{eq:lowT_drift}
\end{equation}
where \(\tau_{eh}(T)\) is the electron--hole scattering time,
\(\tau\) is the momentum-relaxation time due to disorder and/or
phonons, and \(m^*(T)\) is the thermal effective mass of the Dirac
carriers, derived below in Eq.~(\ref{eq:mT}).

Equation~(\ref{eq:lowT_drift}) represents the Dirac-electron and
Dirac-hole sectors by drifting distributions with velocities \(u_e\)
and \(u_h\). This two-drift-velocity closure assumes that intraband
equilibration is fast compared with relaxation of the relative
electron--hole drift,
\(\tau_{ee},\tau_{hh}\ll\tau_{eh}\).
We use it here as a phenomenological description, in the same spirit
as Ref.~\cite{xin}, rather than as a controlled microscopic
approximation. In the nearly symmetric low-temperature Dirac plasma,
the intraband and interband Coulomb times are not parametrically
separated, so Eq.~(\ref{eq:lowT_drift}) should be regarded as the
leading-order reference description of this regime.

Introducing the total and relative velocities,
\(u=u_e+u_h\) and \(v=u_e-u_h\), one finds in the symmetric case
\(u=0\) and
\[
u_e=-u_h=\frac{v}{2}
=-\frac{e\tau_0(T)}{m^*(T)}E,
\]
where
\[
\frac{1}{\tau_0(T)}
=
\frac{1}{\tau_{eh}(T)}
+
\frac{1}{\tau}.
\]
Since \(n_d=p_d\) in this limit, the corresponding resistivity is
\begin{equation}
\rho(T)
=
\frac{m^*(T)}
{e^2[n_d(T)+p_d(T)]\tau_0(T)}
=
\frac{m^*(T)}
{2e^2 n_d(T)\tau_0(T)}.
\label{eq:rho_lowT}
\end{equation}

In the ideal symmetric limit the chemical potential lies at the Dirac
point, \(\mu\simeq0\), and the thermally excited Dirac electrons and
holes have equal densities,
\[
n_d(T)=p_d(T).
\]
Using the Dirac density of states introduced in Sec.~I,
\[
D_d(E)=\frac{g_d|E|}{2\pi(\hbar v_F)^2},
\]
their density is
\begin{equation}
n_d(T)=p_d(T)
=
\int_{0}^{+\infty}
\frac{D_d(E)\,dE}
{\exp\!\left(E/k_BT\right)+1}
=
\frac{g_d\pi}{24}
\frac{(k_BT)^2}{\hbar^2v_F^2}.
\label{eq:dirac_density}
\end{equation}
Here we used
\[
\int_0^\infty\frac{x\,dx}{e^x+1}
=
\frac{\pi^2}{12}.
\]

The thermal effective mass of the neutral Dirac plasma can be defined
by writing its conductivity in Drude form,
\begin{equation}
\sigma
=
\frac{[n_d(T)+p_d(T)]e^2\tau}{m^*(T)}
=
\frac{2n_d(T)e^2\tau}{m^*(T)}.
\label{eq:drude_def}
\end{equation}
The same conductivity obtained directly from the kinetic equation is
\begin{equation}
\sigma
=
2e^2\tau
\int_0^\infty
D_d(E)\frac{v_F^2}{2}
\left(-\frac{\partial f}{\partial E}\right)dE
=
\frac{g_d e^2\tau k_BT\ln2}{2\pi\hbar^2}.
\label{eq:sigma_direct}
\end{equation}
Using
\[
\int_0^\infty
\frac{x e^x\,dx}{(e^x+1)^2}
=
\ln2,
\]
and equating Eqs.~(\ref{eq:drude_def}) and
(\ref{eq:sigma_direct}), we obtain
\begin{equation}
m^*(T)
=
\frac{\pi^2}{6\ln2}
\frac{k_BT}{v_F^2}.
\label{eq:mT}
\end{equation}

If electron--hole scattering dominates the momentum relaxation and
its characteristic rate is taken in the Planckian form
\cite{phillips},
\[
\frac{1}{\tau_{eh}(T)}
=
\frac{k_BT}{h},
\]
then \(\tau_0(T)\simeq\tau_{eh}(T)\propto T^{-1}\).
Since \(m^*(T)\propto T\) and
\(n_d(T)=p_d(T)\propto T^2\),
Eq.~(\ref{eq:rho_lowT}) gives a temperature-independent resistivity,
\begin{equation}
\rho_{\rm CNP}
\simeq
\frac{1}{2\pi g_d\ln2}
\left(\frac{h}{e^2}\right).
\label{eq:rho_const}
\end{equation}
For the single Dirac cone in the present HgTe quantum well,
$g_d=2$, giving $\rho_{\rm CNP}
\simeq
\frac{1}{4\pi\ln2}
\frac{h}{e^2}
\simeq
0.115\,\frac{h}{e^2}.$
The significance of this result is primarily the saturation of the
resistivity rather than the precise numerical prefactor, which depends
on the microscopic electron--hole scattering rate. An analogous
temperature-independent resistivity associated with Planckian
electron--hole scattering has been observed in charge-neutral graphene
\cite{xin}.

The measured low-temperature resistance maximum additionally contains
the disorder-limited percolation contribution discussed in Sec.~III.
Accordingly, the experimental values
$\rho_{\max}=0.25\,h/e^2$ for sample A and
$0.36\,h/e^2$ for sample B exceed the ideal two-fluid estimate,
with the larger value for sample B indicating a stronger contribution
from the disorder/percolation channel.

\subsection{Crossover to a degenerate Dirac-electron--Boltzmann-hole plasma}

As the temperature increases and the heavy holes become thermally
populated, the transport near the charge-neutrality point is no longer
described by the symmetric two-fluid picture of Sec.~IV\,A. The essential
feature of this regime, which we treat following the kinetic theory of
Ref.~\cite{kovalev2}, is dictated by the charge-neutrality condition,
Eq.~(\ref{eq:neutrality_g}). In the high-temperature range, where the Dirac holes become subdominant (\(p_d\to0\)), this condition reduces to the hybrid balance
$n_d\simeq p_{HH}$. Because the heavy-hole density of states greatly exceeds that of the
Dirac carriers, satisfying this balance forces the chemical potential to
rise approximately linearly with temperature. As a result---and this is the central point of the
present description---the Dirac electrons become \emph{moderately
degenerate}, while the heavy holes remain \emph{deeply non-degenerate}.
From the self-consistent solution of Eq.~(\ref{eq:neutrality_g})
[Fig.~1(c)], the electron degeneracy factor is
\(\mu/k_B T\simeq 1.3\text{--}2.2\) over the measured interval, whereas
the heavy-hole activation factor obeys \((\Delta+\mu)/k_B T\gg1\), so the
holes follow Boltzmann statistics. This is opposite to the naive
expectation that all carriers become classical at elevated temperature:
the large heavy-hole density of states pins the Dirac electrons in the
degenerate regime.

For an ideal sharp heavy-hole band edge, the chemical potential  admits the closed-form
Lambert-$W$ solution of Ref.~\cite{kovalev2}, and the moderately degenerate regime
sets in only at $T\gtrsim50$~K. In our samples, however, the unavoidable broadening
of the heavy-hole band edge ($\Gamma_{HH}\sim8$~meV, from well-width fluctuations,
Sec.~I) populates the heavy holes at lower temperatures, so that the
degenerate-electron regime is already established for $T\gtrsim10$~K. We therefore use
the self-consistent densities of Fig.~1(c), which include this broadening, rather than
the sharp-gap asymptotics. The broadening that produces the percolation background of
Sec.~III thus also brings the degenerate hybrid regime down into the experimentally
accessible temperature window.
In this regime the interparticle transport can be treated
\emph{non-perturbatively}. The key simplification, established in Ref.~\cite{kovalev2},
is that the scattering of Dirac electrons by heavy holes is nearly elastic. The
momentum transferred in a collision is of order $q\sim k_BT/(\hbar v_F)$, whereas the
heavy-hole momentum is $p\sim\sqrt{2m_{HH}k_BT}$, so that
\begin{equation}
\frac{q}{p}
\sim
\sqrt{\frac{k_BT}{2m_{\rm HH}v_F^2}}
\ll1,
\qquad
\frac{m_{\rm HH}v_F^2}{k_B}
\simeq5\times10^3~{\rm K}.
\label{eq:quasi_elastic}
\end{equation}

The small recoil can also be seen directly by comparing the effective
Dirac-electron mass at the Fermi level,
$m_F=\mu/v_F^2\sim10^{-3}m_0$, with
$m_{HH}=0.15m_0$. Since $m_F\ll m_{HH}$, the heavy holes recoil only
weakly and act as nearly static scatterers.
This
quasi-elasticity permits a momentum-relaxation time $\tau_{eh}(\varepsilon_k)$ to be
defined for the Dirac electrons, converting the interparticle problem into a
Drude-like one --- in contrast to the first-order perturbative correction employed in
the earlier version of this analysis, which is not controlled once the correction
becomes comparable to the background.

Following Ref.~\cite{kovalev2}, the Dirac-electron conductivity then
takes the Drude form, evaluated at the Fermi energy $\varepsilon_k=\mu$,
\begin{equation}
\sigma_e=\frac{g_d\,e^2}{4\pi\hbar}\,\frac{\mu(T)}{\hbar}
\left[\frac{\tau(\varepsilon_k)\,\tau_{eh}(\varepsilon_k)}
{\tau(\varepsilon_k)+\tau_{eh}(\varepsilon_k)}\right]_{\varepsilon_k=\mu(T)},
\label{eq:sigma_e_drude}
\end{equation}
where $\tau$ is the external (impurity/phonon) momentum-relaxation
time and $\tau_{eh}$ the electron--heavy-hole time. The direct heavy-hole Drude
conductivity is separately negligible owing to the large mass $m_{HH}$: with
$\tau_{HH}\sim10^{-13}$~s it remains below $\sim1.5\%$ of the background over the
fitting interval, despite $p_{HH}\simeq n_d$. Hole--hole collisions do not contribute
to the resistivity, since the parabolic heavy-hole subsystem is Galilean invariant.

\subsection{Interaction model and the \texorpdfstring{$T^2$} resistivity}

The temperature dependence of the resistivity follows from that of
the heavy-hole density and of the electron--heavy-hole relaxation time. The latter is
proportional to the heavy-hole density and depends on the assumed interaction
potential $U_q$. For a short-range contact potential \(U_q=U_0\), used here as a
phenomenological local interaction channel, and for the bare Coulomb
potential \(U_q=2\pi e^2/\epsilon q\), one finds, respectively,\cite{kovalev2},
\begin{equation}
\begin{aligned}
\frac{1}{\tau_{eh}(\varepsilon_k)}
&\propto p_{HH}(T)\,U_0^2\,\varepsilon_k
\qquad (\text{short range}),\\
\frac{1}{\tau_{eh}(\varepsilon_k)}
&\propto \frac{p_{HH}(T)}{\varepsilon_k}
\qquad (\text{Coulomb}).
\end{aligned}
\label{eq:tau_eh_models}
\end{equation}
The heavy-hole density itself is fixed by neutrality. In the
degenerate-electron/Boltzmann-hole regime it grows approximately
quadratically over the experimental fitting window,
\(p_{HH}(T)\propto T^2\), but this is not a universal asymptotic law.
It results from the self-consistent chemical potential together with the
broadened heavy-hole band edge. This approximate scaling is not imposed
in the fit; it is checked directly from the calculated densities in
Sec.~V.
Two limiting behaviors follow. When electron--heavy-hole scattering
dominates over impurity scattering ($N_i<p_{HH}$, the ``clean'' regime), the
conductivity is controlled by $\tau_{eh}$. For the short-range interaction,
$\sigma_e\propto\mu\,\tau_{eh}\propto1/(p_{HH}U_0^2)\propto T^{-2}$, so that the
resistivity rises as $\rho\propto T^2$. In the opposite, impurity-dominated limit
($N_i>p_{HH}$, ``dirty''), $\sigma_e$ is nearly temperature independent. The
long-range Coulomb interaction gives instead $1/\tau_{eh}\propto1/\varepsilon_k$ and
hence $\sigma_e\propto\mu^2/p_{HH}\propto\mathrm{const}$, with no quadratic growth.
The observed quadratic increase of the resistivity therefore selects the short-range
 interaction in the clean regime.

It is convenient to separate the inverse conductivity into a
temperature-independent disorder background and an interaction-induced
contribution. For the short-range interaction in the clean regime, this
gives the compact form
\begin{equation}
\begin{aligned}
\frac{1}{s(T)} &=
\frac{1}{s_0}
+\frac{1}{g_d}\left(\frac{m_{HH}U_0}{\hbar^2}\right)^2 X_{HH}(T),\\
X_{HH}(T) &=
\frac{\pi\hbar^2\,p_{HH}(T)}{m_{HH}^2v_F^2}.
\end{aligned}
\label{eq:fit_form}
\end{equation}
where
$s=\sigma/(e^2/\hbar)$ and
$s_0=\sigma_0/(e^2/\hbar)$, with $\sigma_0$ denoting the
temperature-independent disorder background, while $p_{\rm HH}(T)$
is the total heavy-hole sheet density obtained from the self-consistent
calculation. The factor $1/g_d$ does not enter the charge-neutrality condition;
it appears only when converting the per-Dirac-flavor expression of
Ref.~[18] to the total conductivity. Writing Eq.~(16) in resistivity
form gives 
\begin{equation}
\begin{aligned}
\rho(T) &= \rho_0+\rho_1\,X_{HH}(T),\\
\rho_1 &= \frac{m_{HH}^2\,U_0^2}{g_d\,e^2\hbar^3}.
\end{aligned}
\label{eq:rho_fit}
\end{equation}
so that the excess resistivity is proportional to the
self-consistently computed heavy-hole density, $\Delta\rho\propto p_{HH}(T)$. The
impurity/broadening parameters entering $\sigma_0$ cancel from the interaction
coefficient $\rho_1$, so that the short-range amplitude $U_0$ can be extracted from
the slope alone, independently of the disorder background. The quantitative comparison
and the extraction of $U_0$ are presented in Sec.~V.

Finally, we note that for the broadening relevant here
($\Gamma_{HH}\sim8$~meV) the heavy holes are already appreciable at $T\sim10$~K while
the Dirac-hole density is subdominant, so the genuine three-component regime, in which
Dirac electrons, Dirac holes, and heavy holes contribute simultaneously, occupies a
narrow temperature interval and is difficult to resolve experimentally. For
$T\gtrsim10$~K the system is well described as the degenerate two-component
Dirac-electron--Boltzmann-hole plasma treated above.

\section{Quantitative analysis of the temperature dependence}

We now compare the measured temperature dependence with the
degenerate Dirac-electron--Boltzmann-hole model of Sec.~IV.
For $T\gtrsim10$~K, the Dirac-hole density is subdominant
[Fig.~1(c)], the Dirac electrons are moderately degenerate
($\mu/k_BT\simeq1.3$--$2.2$), and the heavy holes obey Boltzmann
statistics. In this regime, electron--heavy-hole scattering is
quasi-elastic and admits a relaxation-time description, so the
transport is treated non-perturbatively rather than as a first-order
interaction correction. For a short-range interaction in the clean
limit, the model yields $\sigma_e\propto T^{-2}$ and hence an
approximately quadratic resistivity, whereas a long-range Coulomb
interaction gives a temperature-independent conductivity.

The comparison uses the two-parameter form of
Eq.~(\ref{eq:rho_fit}),
\[
\rho(T)=\rho_0+\rho_1 X_{HH}(T),
\]
so that the excess resistivity is proportional to the heavy-hole
density,
\[
\Delta\rho(T)\propto p_{HH}(T).
\]
The temperature-dependent density $p_{HH}(T)$ is not an adjustable
parameter but is taken directly from the self-consistent calculation
of Fig.~1(c), with $\Gamma_{HH}=8$~meV, $g_d=2$, and $g_{HH}=4$.
The only fit parameters are the constant background $\rho_0$ and the
interaction amplitude $U_0$. The impurity parameters that determine
the background conductivity $\sigma_0$ do not enter the interaction
coefficient $\rho_1$, so that $U_0$ is determined by the
temperature-dependent part of the resistivity.

\begin{figure}[ht]
\centering
\includegraphics[width=8.5cm]{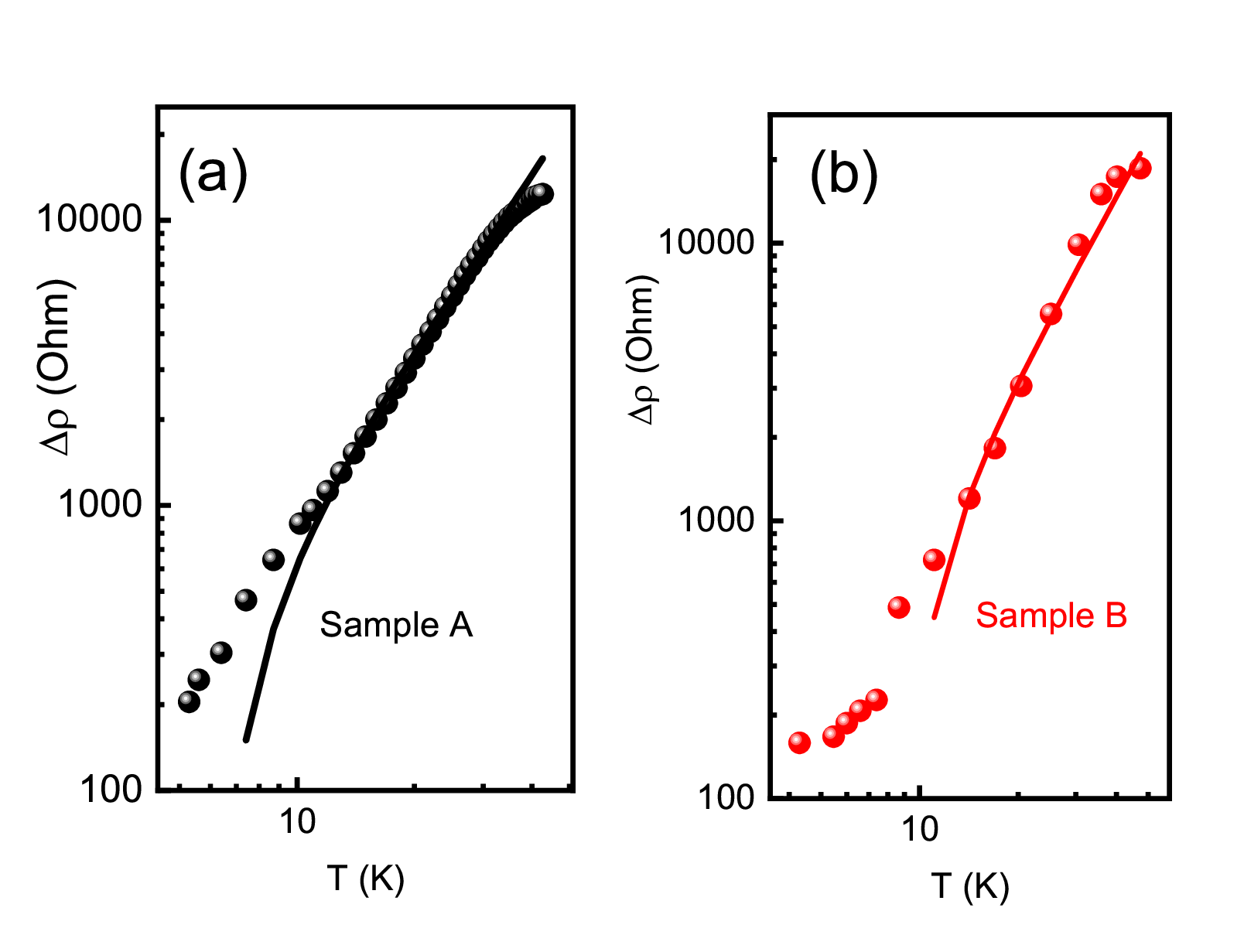}
\caption{\label{fig:rho}
(Color online) Excess resistivity $\rho(T)-\rho_0$ at the fixed
low-temperature reference voltage $V_0$ for samples A (a) and B (b).
The solid line is the non-perturbative model of Sec.~IV,
$\rho=\rho_0+\rho_1X_{\rm HH}(T)$, with the heavy-hole density taken
from the calculation of Fig.~1(c)}
\end{figure}

Figure~\ref{fig:rho} shows the comparison for both samples.
For sample A, measured on a dense and uniform temperature grid,
a free-exponent fit
$\Delta\rho\propto T^p$ over $10$--$30$~K gives
$p=2.12\pm0.02$ ($R^2=0.998$). The local logarithmic slope remains
in the range $2.1$--$2.3$ and decreases above approximately
$30$~K, showing that the dependence is approximately quadratic
rather than an exact power law. The residuals of the model fit
are small for sample A (rms $\sim6\%$).

Sample B was measured on a sparser temperature grid and, being
the more resistive device
($\rho_{\max}=0.36$ versus $0.25\,h/e^2$), is more affected by
self-heating. It therefore shows larger scatter, with residuals up
to approximately $20\%$, and a nominally steeper exponent
$p\simeq2.6$. We consequently use sample A for the quantitative
analysis, while sample B provides a qualitative consistency check
that the resistance increases in the same temperature range in
which the heavy-hole population becomes important.

\begin{figure}[ht]
\centering
\includegraphics[width=8.5cm]{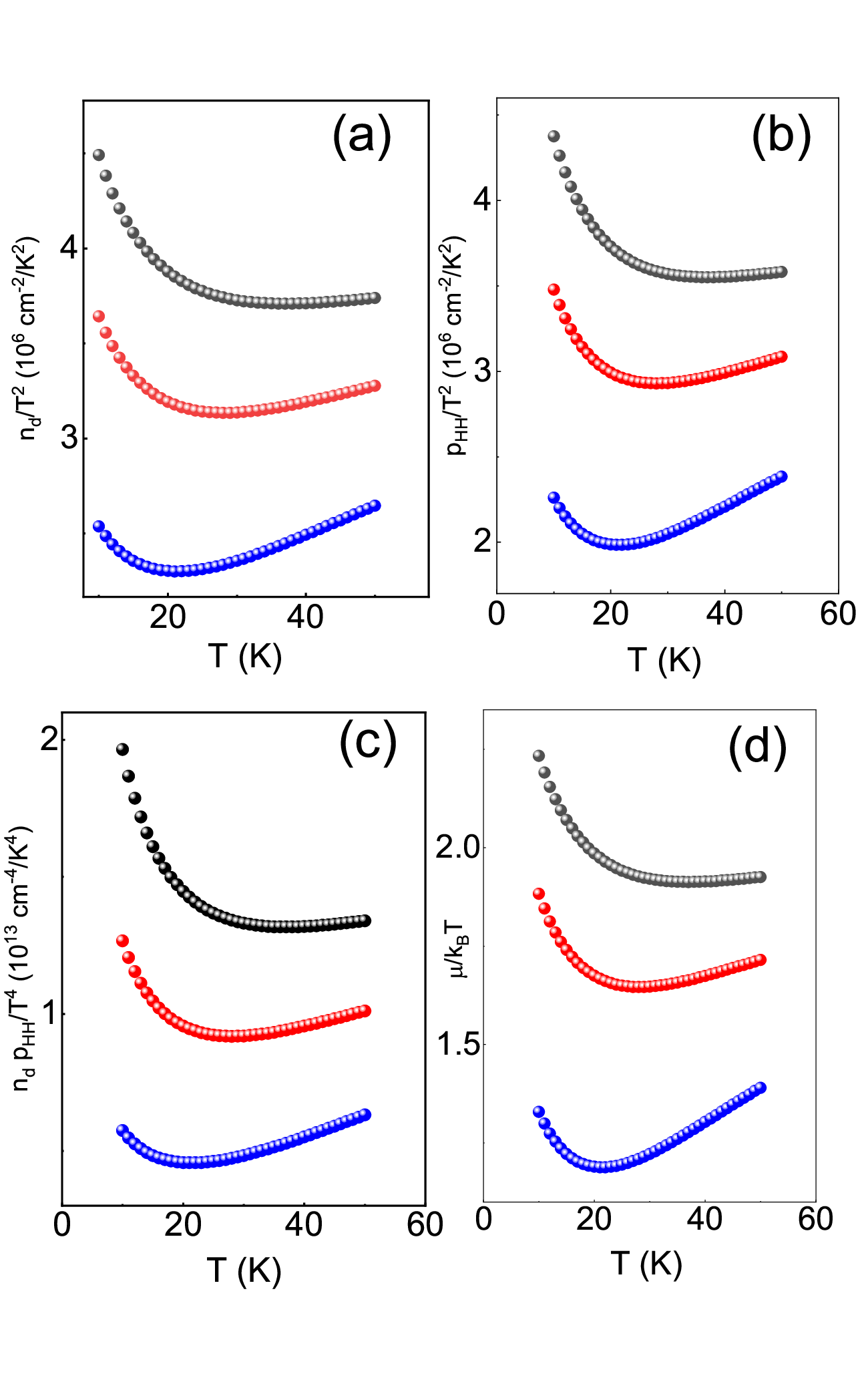}
\caption{\label{fig:densitydiag}
(Color online) Density diagnostics with the chemical potential
$\mu(T)$ determined self-consistently from charge neutrality,
$n_d=p_d+p_{HH}$, for $\Gamma_{HH}=8$~meV and
$\Delta=15$, $17$, and $20$~meV.
Panels show (a) $n_d/T^2$, (b) $p_{HH}/T^2$,
(c) $n_dp_{HH}/T^4$, and (d) $\mu/k_BT$.
The finite temperature dependence of these normalized quantities
shows that the approximately quadratic carrier-density dependence
is not an exact scaling law but emerges from the self-consistent
evolution of the multicomponent carrier populations.}

\end{figure}

Because the system is not in the asymptotic limit
$k_BT\gg\Delta$
($\Delta/k_B\simeq175$~K, whereas the measurements extend only to
approximately 47~K), the observed near-quadratic
behavior should not be interpreted as a universal high-temperature
scaling law. To clarify this point, Fig.~\ref{fig:densitydiag}
shows the normalized quantities $n_d/T^2$, $p_{HH}/T^2$,
$n_dp_{HH}/T^4$, and the degeneracy parameter $\mu/k_BT$.
If the carrier densities followed exact quadratic laws, the first
three quantities would be temperature independent. Instead, the
calculated curves retain a finite and smooth temperature dependence.
Thus, the near-$T^2$ behavior observed over the experimental fitting
interval is an approximate numerical property of the self-consistent
carrier populations rather than an exact asymptotic scaling law.

Fitting Eq.~(\ref{eq:rho_fit}) over $10$--$30$~K gives
$\rho_0=6.3~\mathrm{k\Omega}$ and
$U_0=93~\mathrm{eV\,nm^2}$ ($R^2=0.998$) for sample A, and
$\rho_0=9.3~\mathrm{k\Omega}$ and
$U_0=95~\mathrm{eV\,nm^2}$ ($R^2=0.982$) for sample B.
The two samples therefore yield essentially the same effective
interaction amplitude,
\[
U_0\simeq90~\mathrm{eV\,nm^2},
\]
while the disorder-dependent background $\rho_0$ differs between
them.

This magnitude is of the natural Coulomb scale for the low carrier
density realized near charge neutrality. With
$N_e\simeq(1.5$--$6.5)\times10^9$~cm$^{-2}$ over the fitting
interval, the Fermi momentum is
\[
k_F=\sqrt{2\pi N_e}
\simeq(1.0--2.0)\times10^7~\mathrm{m^{-1}},
\]
and the two-dimensional Thomas--Fermi screening wave vector satisfies
\[
\frac{q_{TF}}{k_F}
=
g_d\alpha
\simeq0.3--0.6,
\]
with
\[
\alpha=\frac{e^2}{\epsilon\hbar v_F}
\simeq0.16--0.31
\]
for $\epsilon=10--20$. Since $q_{TF}<k_F$, screening is weak at
these densities, and the relevant interaction is close to the bare
Coulomb value at the Fermi momentum,
\[
V(k_F)=\frac{2\pi e^2}{\epsilon k_F}
\simeq40--90~\mathrm{eV\,nm^2}.
\]
The extracted $U_0$ is therefore of the expected order.
It is much larger than the strongly screened estimate
\[
U_0^{\rm ov}
=
\frac{2\pi\hbar^2}{g_v m_{HH}}
\simeq1.6~\mathrm{eV\,nm^2},
\]
which assumes $q_{TF}\gg k_F$ and is not appropriate for the dilute
charge-neutrality regime considered here.

We stress that $U_0$ remains an effective interaction parameter.
The observed near-$T^2$ dependence is reproduced by the contact
form of the interaction, whereas a fully microscopic calculation
would require the momentum dependence of the screened interaction
and the finite width of the quantum well, including penetration of
the carrier wave functions into the barriers. These effects are not
included in the contact-interaction model of Ref.~\cite{kovalev2}.
The extracted value should therefore be regarded as an
order-of-magnitude interaction scale rather than a microscopically
converged amplitude. The degeneracy factors $g_d$ and $g_{HH}$ are
fixed band-structure parameters and are not adjustable quantities.

Unlike a first-order interaction correction, the present
description does not rely on the interaction contribution being
small compared with the background resistivity. The relaxation-time
treatment is justified by the quasi-elasticity of the
electron--heavy-hole scattering: because
$m_F/m_{HH}\sim10^{-2}$, the recoil of the heavy holes is small.

The deviations between the model and experiment at the highest
temperatures in Fig.~\ref{fig:rho} may arise from effects neglected
in the minimal treatment, including direct heavy-hole Drude
conductivity, phonon scattering, the departure of the carrier
densities from their approximate quadratic temperature dependence,
and the fact that the electron degeneracy is only moderate
($\mu/k_BT\sim2$). We therefore use the $10$--$30$~K interval for
the quantitative determination of the approximately quadratic
temperature dependence.

The observation of a near-$T^2$ resistivity remains informative.
Within the degenerate hybrid picture, it is consistent with
short-range friction between moderately degenerate Dirac electrons
and thermally populated Boltzmann heavy holes, with the temperature
dependence determined by the self-consistent heavy-hole population.
This provides a quantitative constraint for microscopic theories of
multicomponent Dirac plasmas in HgTe quantum wells and their
crossover from the low-temperature Dirac plasma to the
Dirac-electron--heavy-hole regime.
\section{Conclusion}

We have studied charge transport near the charge-neutrality point in near-critical HgTe
quantum wells hosting Dirac electrons, Dirac holes, and heavy holes from lateral
valence-band valleys. The self-consistent carrier densities, obtained
from the charge-neutrality condition with the spin/valley degeneracies restored,
reveal a crossover from a low-temperature symmetric two-component Dirac plasma to a
high-temperature hybrid plasma. A central feature of the latter, dictated by charge
neutrality, is that the large heavy-hole density of states pins the chemical potential
above the Dirac point: the Dirac electrons remain \emph{moderately degenerate}
($\mu/k_BT\simeq1.3$--$2.2$ over the measured range), while the thermally activated
heavy holes stay \emph{deeply non-degenerate} and obey Boltzmann statistics.

Experimentally, we find that the resistance increases with temperature and follows an
approximately \(T^2\) dependence above $\approx10$~K (fitted exponent
$p=2.12\pm0.02$ for the higher-quality sample). This behavior differs from the
temperature-independent resistance expected for a symmetric Dirac electron--hole
plasma. We attribute it to the thermal population of heavy-hole states, which
progressively replace the Dirac holes and provide an additional
interaction-induced current-relaxation channel.

Because the electrons are degenerate while the heavy holes are heavy
and non-degenerate, the electron--heavy-hole scattering is nearly elastic
($m_F/m_{HH}\sim10^{-2}$), which allows the interparticle transport to be described by
a momentum-relaxation time rather than by a first-order perturbative correction.
Within this non-perturbative treatment, following Ref.~\cite{kovalev2}, a short-range
 interaction in the clean limit yields a Dirac-electron
conductivity $\sigma_e\propto T^{-2}$ and hence a quadratic resistivity, whereas a
long-range Coulomb interaction gives a temperature-independent conductivity. The
observed quadratic increase therefore selects the short-range channel. The excess
resistivity is proportional to the self-consistently computed heavy-hole density,
$\Delta\rho\propto p_{HH}(T)$, and is well described by a two-parameter expression in
which only the disorder-limited background and the interaction amplitude are adjusted;
the carrier densities are taken from the self-consistent
reference-spectrum calculation and are not additional fitting parameters.

The extracted short-range amplitude, $U_0\simeq90$~eV\,nm$^2$, is of
the order of the bare Coulomb interaction evaluated at the Fermi momentum for the
dilute charge-neutrality density, $V(k_F)=2\pi e^2/\epsilon k_F$, consistent with the
weak screening ($q_{TF}<k_F$) realized in this regime. It should be understood as an
effective parameter, since the contact form of the interaction and the finite width of
the quantum well are treated only approximately.

We also find that the moderately degenerate hybrid regime, which for
an ideal sharp gap would set in only above $\approx50$~K, is brought down into the
experimentally accessible window ($T\gtrsim10$~K) by the broadening of the heavy-hole
band edge caused by well-width fluctuations --- the same disorder that produces the
percolation background near charge neutrality. A complete microscopic description of
the low- and intermediate-temperature crossover, where the disorder-induced
percolation channel, Dirac-hole transport, and interparticle scattering contribute
simultaneously, remains challenging and is left for future work. A decisive test of this interpretation would require temperature- and
gate-dependent Hall, magnetoresistance, and thermal-transport measurements.
However, in the present hybrid regime such data cannot be analyzed within a
simple single-carrier Drude picture. This is already illustrated by related
Hall measurements in HgTe quantum wells in the valence-band regime, where
Dirac holes and heavy holes coexist. In that case the Hall response was found
to be strongly anomalous and could exceed the ordinary single-band Hall value
by about an order of magnitude~\cite{hall2025}. This reflects the
multicomponent nature of the transport and the importance of intersubband
scattering. The charge-neutral regime studied here is even more involved,
because Dirac electrons, Dirac holes, and heavy holes may contribute
simultaneously, with different mobilities and intercomponent friction. A
systematic Hall and magnetotransport analysis of this three-component regime
therefore requires a separate kinetic treatment and will be addressed in
future work.

Our results establish near-critical HgTe quantum wells as a versatile platform for
exploring interaction-driven transport in multicomponent Dirac systems,
in which the coexistence of degenerate massless and non-degenerate
massive carriers, tunable through temperature and gate voltage, gives direct access to quantum friction between distinct fermionic species, enabled by the spectral non-invariance of the Dirac dispersion that unlocks the current from the total momentum.
\section {ACKNOWLEDGMENTS}
This work is supported by FAPESP (São Paulo Research Foundation) Grants No. 2019/16736-2, No. 2021/12470- 8, No. 2025/22786-3, CNPq (National Council for Scientific and Technological Development). The growth of HgTe quantum wells  was supported by the Russian Science Foundation (Grant No. 23-72-30003)
\section{DATA AVAILABILITY}
The data that support the findings of this article are openly
available \cite{zenodo}.


\begin{thebibliography}{32}

\bibitem{pal} H. K. Pal, V. I. Yudson, and D. L. Maslov, Resistivity of non
Galilean invariant term Fermi and non Fermi liquids, Lith. J. Phys. {\bf 52}, 142 (2012).
\bibitem{nam}
Y. Nam, D.-K. Ki, D. Soler-Delgado, A. F. Morpurgo, Electron-hole collision limited 
transport in charge-neutral bilayer graphene. Nat. Phys. {\bf 13}, 1207 (2017).
\bibitem{tan}
Cheng Tan, Derek Y. H. Ho, Lei Wang, Jia I. A. Li, Indra Yudhistira, Daniel A. Rhodes,
Takashi Taniguchi, Kenji Watanabe, Kenneth Shepard, Paul L. McEuen, Cory R. Dean,
Shaffique Adam, James Hone, Dissipation-enabled hydrodynamic conductivity in a tunable bandgap semiconductor,  Sci. Adv. {\bf 8}, 8481 (2022).
\bibitem{olshanetsky}
E. B. Olshanetsky, Z. D. Kvon, M. V. Entin, L. I. Magarill, N. N. Mikhailov, and
S. A. Dvoretsky. Scattering processes in a two-dimensional semimetal, JETP Lett.  {\bf 89}, 290 (2009).
\bibitem{entin}
M. V. Entin, L. I. Magarill, E. B. Olshanetsky, Z. D. Kvon, N. N. Mikhailov, and
S. A. Dvoretsky, The effect of electron-hole scattering on transport properties of a 2D semimetal in the HgTe quantum well. JETP,  {\bf 117}, 933 (2013).

\bibitem{hwang}
E. H. Hwang and S. Das Sarma, Temperature dependent resistivity of spin-split subbands in GaAs two-dimensional hole systems, Phys. Rev. B {\bf 67}, 115316 (2003).

\bibitem{nagaev1}
K. E. Nagaev, and A. A. Manoshin, Electron-electron scattering and transport properties of spin-orbit coupled electron gas,
Phys. Rev. B {\bf 102}, 155411 (2020).

\bibitem{nagaev2}
K. E. Nagaev, Electron-electron scattering and conductivity of disordered systems with a Galilean-invariant spectrum, Phys. Rev. B {\bf 106}, 085411 (2022).
\bibitem{gusev}
G. M. Gusev,  A. D. Levin, E. B. Olshanetsky, Z. D. Kvon, V. M. Kovalev, M. V. Entin,  and N. N. Mikhailov, Interaction-dominated transport in two-dimensional conductors: From degenerate to partially degenerate regime, Phys.Rev.B, {\bf 109}, 035302 (2024).
\bibitem{levin}
A. D. Levin, G. M. Gusev, F. G. G. Hernandez, E. B. Olshanetsky, V. M. Kovalev, M. V. Entin,  and N. N. Mikhailov, Interaction-controlled transport in a two-dimensional massless-massive Dirac system: Transition from degenerate to nondegenerate regimes, Phys.Rev.Research, {\bf 6}, 023121 (2024).

\bibitem{kovalev}
V. M. Kovalev, M. V. Entin, Z. D. Kvon,  A. D. Levin, V. A. Chitta, G. M. Gusev,  and  N. N. Mikhailov, Resistivity of non-Galilean invariant two dimensional Dirac system, Phys.Rev.Lett., {\bf 134}, 196303 (2025).
\bibitem{snegirev}
A. V. Snegirev , V. M. Kovalev , M. V. Entin , E. B. Olshanetsky, G. M. Gusev ,
N. N. Mikhailov, Interplay between electron-hole and impurity scattering in HgTe-based quantum wells, Phys.Rev.Research, {\bf 7}, 043169 (2025)

\bibitem{xin}
 Na Xin, James Lourembam, Piranavan Kumaravadivel, A. E. Kazantsev, Zefei Wu, 
Ciaran Mullan, Julien Barrier, A. A. Geim, I. V. Grigorieva, A. Mishchenko, 
A. Principi, V. I. Fal’ko1, L. A. Ponomarenko, A. K. Geim, and Alexey I. Berdyugin,
Giant magnetoresistance of Dirac plasma in high-mobility graphene, Nature {\bf 616}, 270–274 (2023).
\bibitem{sarma}
S. Das Sarma, Shaffique Adam, E. H. Hwang, and Enrico Rossi, Electronic transport in two-dimensional graphene, Rev. Mod. Phys. {\bf 83}, 407, (2011).

\bibitem{crossno2016}
J. Crossno, J. K. Shi, K. Wang, X. Liu, A. Harzheim, A. Lucas, S. Sachdev, P. Kim,
T. Taniguchi, K. Watanabe, T. A. Ohki, and K. C. Fong,
Observation of the Dirac fluid and the breakdown of the Wiedemann--Franz law in graphene,
\textit{Science} \textbf{351}, 1058--1061 (2016).

\bibitem{ponomarenko}
L. A. Ponomarenko and A. Principi and A. D. Niblett and W. Wang and R. V. Gorbachev and P. Kumaravadivel and A. I. Berdyugin and A. V. Ermakov and S. Slizovskiy and K. Watanabe and T. Taniguchi and Q. Ge and V. I. Fal'ko and L. Eaves and M. T. Greenaway and A. K. Geim, Extreme electron--hole drag and negative mobility in the {Dirac} plasma of graphene, Nat. Commun., {\bf 15}, 9869, (2024).
\bibitem{phillips}
Phillips, P. W., Hussey, N. E. and Abbamonte, P. Stranger than metals. Science { \bf 377},
4273 (2022).

\bibitem{kovalev2}
Y. Huang and O. V. Kibis and V. M. Kovalev and I. G. Savenko, Conductivity of charge-neutral multicomponent 2D electron-hole system, arXiv:2604.07846,
(2026).
\bibitem{raichev2012}
O. E. Raichev, Effective Hamiltonian, energy spectrum, and phase transition induced by in-plane magnetic field in symmetric HgTe quantum well, Phys. Rev. B {\bf 85},  045310 (2012).
\bibitem{kristopenko2016}
S. S. Krishtopenko, I. Yahniuk, D. B. But, V. I. Gavrilenko, W. Knap, and F. Teppe,
Phys. Rev. B {\bf 94}, 245402 (2016).
\bibitem{buttner}
B. Büttner, C. X. Liu, G. Tkachov, E. G. Novik, C. Brne, H. Buhmann, E. M. Hankiewicz, P. Recher, B. Trauzettel, S. C. Zhang, and L. W. Molenkamp, Single valley Dirac fermions in
zero-gap HgTe quantum wells, Nat. Phys. {\bf 7}, 418 (2011).
\bibitem{kozlov2013}
D. A. Kozlov, Z. D. Kvon, N. N. Mikhailov, and S. A.Dvoretskii, Weak localization of Dirac fermions in HgTe quantum wells, JETP Lett. {\bf96}, 730 (2013).
\bibitem{gusev2}
G. M. Gusev, D. A. Kozlov, A. D. Levin, Z. D. Kvon, N. N. Mikhailov, and S. A. Dvoretsky, Robust helical edge transport at $\nu = 0$ quantum Hall state, Phys. Rev. B {\bf 96}, 045304 (2017).
 \bibitem{kristopenko2019}
 S. S. Krishtopenko, W. Desrat, K. E. Spirin, C. Consejo, S. Ruffenach, F. Gonzalez-Posada, B. Jouault, W. Knap, K. V.Maremyanin, V. I. Gavrilenko, G. Boissier, J. Torres, M. Zaknoune, E. Tournié, and F. Teppe, Massless Dirac fermions in III-V semiconductor quantum wells, Phys. Rev. B {\bf 99}, 121405(R) (2019).
\bibitem{yahniuk2019}
I. Yahniuk, S. S. Krishtopenko, G. Grabecki, B. Jouault, C. Consejo, W. Desrat, M. Majewicz, A. M. Kadykov, K. E. Spirin, V. I. Gavrilenko, N. N. Mikhailov, S. A. Dvoretsky, D. B. But, F. Teppe, J. Wróbel, G. Cywiński, S. Kret, T. Dietl, and W. Knap,
Magneto-transport in inverted HgTe quantum wells,
npj Quantum Mater. \textbf{4}, 13 (2019).
\bibitem{gusev2022}
G. M. Gusev, Z. D. Kvon, D. A. Kozlov, E. B. Olshanetsky,M. V. Entin, and N. N. Mikhailov, Transport through the network of topological channels in HgTe based quantum well, 2D Mater. {\bf9}, 015021 (2022).

\bibitem{Wiedmann2015}
S. Wiedmann, A. Jost, C. Thienel, C. Brüne, P. Leubner, H. Buhmann, L. W. Molenkamp, J. C. Maan, and U. Zeitler,
Temperature-driven transition from a semiconductor to a topological insulator,
Phys. Rev. B \textbf{91}, 205311 (2015).

\bibitem{Marcinkiewicz2017}
M. Marcinkiewicz, S. Ruffenach, S. S. Krishtopenko, A. M. Kadykov, C. Consejo, D. B. But, W. Desrat, W. Knap, J. Torres, A. V. Ikonnikov, K. E. Spirin, S. V. Morozov, V. I. Gavrilenko, N. N. Mikhailov, S. A. Dvoretskii, and F. Teppe,
Temperature-driven single-valley Dirac fermions in HgTe quantum wells,
Phys. Rev. B \textbf{96}, 035405 (2017).

\bibitem{Kadykov2018}
A. M. Kadykov, S. S. Krishtopenko, B. Jouault, W. Desrat, W. Knap, S. Ruffenach, C. Consejo, J. Torres, S. V. Morozov, N. N. Mikhailov, S. A. Dvoretskii, and F. Teppe,
Temperature-Induced Topological Phase Transition in HgTe Quantum Wells,
Phys. Rev. Lett. \textbf{120}, 086401 (2018).

\bibitem{kozlov2014}
D. A. Kozlov, Z. D. Kvon, N. N. Mikhailov, S. A. Dvoretskii, S. Weishaupl, 
Y. Krupko, and J.-C. Portal, Appl. Phys. Letter.{\bf 105} , 132102 (2014).
\bibitem{hall2025}
G. M. Gusev, A. D. Levin, V. A. Chitta, Z. D. Kvon and  N. N. Mikhailov, Interaction-induced  magnetotransport in a 2D Dirac–Heavy hole hybrid band system, Scientific Reports {\bf 16},  2381 (2026).
\bibitem{Supplemental}
See Supplemental Material at [URL will be inserted by publisher]
for details of the simplified three-carrier conductivity calculation
used to estimate the temperature-induced shift of the resistance maximum.
\bibitem{zenodo}
 G. M. Gusev, Data set: Transport in HgTe . Zenodo
(2026), doi: https://doi.org/10.5281/zenodo.19822128

\end{thebibliography}
\end{document}